\documentclass[a4paper,11pt]{article}
\pdfoutput=1
\usepackage{jheppub} 

\newcommand\fft[2]{\frac{#1}{#2}}
\newcommand\nn{\nonumber}
\DeclareMathOperator{\Tr}{Tr}
\DeclareMathOperator{\PE}{PE}

\preprint{LITP-26-24}

\title{\boldmath Giant gravitons without giant gravitons}

\author{Blake J. Bottesi}
\emailAdd{bottesib@umich.edu}
\author{and James T. Liu}
\emailAdd{jimliu@umich.edu}
\affiliation{Leinweber Institute for Theoretical Physics, University of Michigan, Ann Arbor, MI 48109-1040, USA}

\abstract{We demonstrate that the partition function of a free bosonic $U(N)$ multi-matrix model admits a giant graviton expansion.  For $d$ matrices, there are $d$ distinct giant gravitons, and the expansion can be characterized by a $d$-node quiver with bifundamentals connecting the nodes.  This suggests the intriguing possibility of a more generalized holography where models without any obvious gravitational dual can nevertheless be given a dual brane interpretation.  The giant graviton expansion of this multi-matrix model is somewhat unconventional though, as the giant graviton partition functions will necessarily have intrinsic $N$-dependence.}

\begin{document}
\maketitle
\flushbottom

\section{Introduction}
\label{sec:intro}

AdS/CFT at finite $N$ exhibits a remarkable amount of structure such as competing saddles, fortuity and erratic behavior at large but finite $N$.  One interesting place where such effects show up is in the BPS partition function of holographic $U(N)$ gauge theories, where trace relations at finite $N$ become important.  These relations first enter at $\mathcal O(N)$ and are intimately tied to the presence of giant gravitons, namely wrapped D3-branes, on the gravity side of the duality.

A fruitful way to examine the role of giant gravitons in finite-$N$ holography is through the giant graviton expansion of the superconformal index.  This index, $I_N(\vec q\,)$, is a refined Witten index that depends on a set of fugacities, $\vec q=\{q_1,q_2,\ldots,q_n\}$, and the rank $N$ of the gauge group.  This index generally takes on a simple form, $I_\infty(\vec q\,)$, in the infinite-$N$ limit, as trace relations drop out and the combinatorics simplify.  The giant graviton expansion then has the form \cite{Bourdier:2015wda,Bourdier:2015sga,Arai:2019xmp,Arai:2019wgv,Arai:2019aou,Arai:2020qaj,Arai:2020uwd,Imamura:2021ytr,Gaiotto:2021xce}
\begin{equation}
    \fft{I_N(\vec q\,)}{I_\infty(\vec q\,)}=1+\sum_{\vec m\ne0}\vec q\;^{\vec m N}\hat I^{\mathrm{GG}}_{(\vec m)}(\vec q\,),
\label{eq:GGE}
\end{equation}
where $\vec m=\{m_1,m_2,\ldots,m_n\}$ is a set of non-negative integers counting the number of giants associated with each fugacity $q_i$.  This expansion has a beautiful interpretation on the gravity side; the infinite-$N$ index $I_\infty(\vec q\,)$ counts the fluctuation of Kaluza-Klein gravitons, while $\hat I^{\mathrm{GG}}_{(\vec m)}(\vec q\,)$ counts the fluctuations on a stack of $\vec m$ giant gravitons.  As each giant carries angular momentum (i.e.~charge) in the internal space, the pre-factor $\vec q\;^{\vec mN}$ accounts for the charge from the classical motion of the stack of giants.

While the giant graviton expansion has a natural holographic interpretation in terms of wrapped D3-branes \cite{Imamura:2021ytr}, Murthy demonstrated in \cite{Murthy:2022ien} that any $U(N)$ matrix model partition function of the form
\begin{equation}
    Z_N(\vec q\,)=\int_{U(N)} dU\PE(f(\vec q\,)\chi_{\mathrm{adj}}(U)),
\end{equation}
where PE denotes the plethystic exponential, admits a similar expansion of the form
\begin{equation}
    \fft{Z_N(\vec q\,)}{Z_\infty(\vec q\,)}=1+\sum_{m>0}G_N^{(m)}(\vec q\,),
\label{eq:Murthy}
\end{equation}
where the leading order behavior of $G_N^{(m)}(\vec q\,)$ is given by $G_N^{(m)}(\vec q\,)\sim |\vec q\,|^{m(N+m)}$, assuming $f(\vec q\,)\sim |\vec q\,|+\cdots$ \cite{Murthy:2022ien,Liu:2022olj}.  This expansion is obtained directly from the matrix integral, and, in contrast to (\ref{eq:GGE}), only includes a single term $G_N^{(m)}(\vec q\,)$ at each order in the expansion.  Moreover, the $N$ dependence of the terms is contained within the $G_N^{(m)}(\vec q\,)$.  While the two expansions are distinct \cite{Liu:2022olj}, it was shown explicitly for the 1/2 BPS index \cite{Eniceicu:2023uvd} and Schur index \cite{Ezroura:2024wmp} of $\mathcal N=4$ SYM that Murthy's expansion, (\ref{eq:Murthy}), can be rearranged to match the wrapped D3-brane expansion, (\ref{eq:GGE}).  This is expected to hold more generally for the 1/16 BPS index as well, although this has yet to be shown conclusively.

Although most applications of Murthy's expansion have been to put the giant graviton expansion of the superconformal index on a firmer mathematical footing, it is noteworthy that this expansion can be applied to \emph{any} matrix partition function of the form (\ref{eq:Murthy}).  In particular, this expansion can be developed even for non-supersymmetric partition functions, such as the thermal partition functions of SYM recently studied in \cite{Budzik:2026vqe}.  This poses the question: how universal are giant graviton expansions, and do they exist even in theories without any obvious dual brane interpretation?  From a holographic point of view, it is natural to associate giant gravitons with wrapped branes.  But presumably not every large-$N$ matrix model has a holographic brane dual.  One can also ask the related question of what single letter partition functions $f(\vec q\,)$ in (\ref{eq:Murthy}) admit a dual brane interpretation.

It is also noteworthy that Murthy's expansion can be applied to completely non-supersymmetric models and not just to non-BPS partition functions of supersymmetric theories.  However, this raises the issue of whether non-supersymmetric and even non-conformal theories can have a holographic dual and a giant graviton expansion.  In order to address these questions, we will investigate the partition function of a purely bosonic multi-matrix model.  This model is that of $d$ Hermitian $N\times N$ matrices with a free action, so the partition function just counts the number of independent gauge invariant operators that can be formed out of these matrices.  While this model was studied as a model of bosonic fortuity in Ref.~\cite{deMelloKoch:2025cec}, we will show that it also admits a brane-like giant graviton expansion of the form (\ref{eq:GGE}), however with one important difference in that some of the giant graviton indices, $\hat I^{\mathrm{GG}}_{(\vec m)}(\vec q\,)$, will necessarily depend on $N$.

Because of this $N$-dependence, the bosonic giant graviton expansion is distinct from that of the D3-brane giants pertaining to $\mathcal N=4$ SYM.  Nevertheless, there are many similarities, with perhaps the most striking one being that there is a special set of `standard' giant graviton configurations that take the form of a quiver field theory with vector and bifundamental contributions to the giant graviton index.  This mirrors the picture of D3-brane giants in AdS$_5\times S^5$ even though the bosonic giants do not have any obvious connection to any physical brane worldvolume theory.

For the case of a single bosonic matrix, $X$, the free partition function counts all states built out of products of traces $\Tr(X^n)$.  As this is identical to the states enumerated by the 1/2 BPS index of $\mathcal N=4$ SYM, the bosonic partition function of the single matrix model is identical to the 1/2 BPS index and hence admits the same giant graviton expansion.  This suggests that the D3-brane interpretation of this index is not essential, and that the giant graviton expansion in the form of (\ref{eq:GGE}), but allowing for $N$-dependent indices, can be developed intrinsically in any theory that can be encoded as a unitary matrix model.  We will provide evidence for this conjecture below by studying the two-matrix model partition function.

This paper is organized as follows.  In section~\ref{sec:gge} we apply Murthy's giant graviton expansion to the bosonic multi-matrix model partition function and show that it can be rearranged into a sum over $\hat I^{\mathrm{GG}}_{(\vec m)}$, as in (\ref{eq:GGE}).  Then, in section~\ref{sec:d=2}, we provide further evidence for a holographic brane interpretation of the bosonic partition function in the case of two-matrix model.  As mentioned above, while the bosonic giant graviton expansion has the same structure as that of the superconformal index, it nevertheless differs in that the giant graviton indices $\hat I^{\mathrm{GG}}_{(\vec m)}$ can in some cases depend on $N$.  We show explicitly how this $N$-dependence arises in section~\ref{sec:Ndepend}.  Finally, we conclude in section~\ref{sec:disc} with some additional thoughts about the $N$-dependence of the giant graviton indices and whether it is restricted to  bosonic models or may show up in the expansion of supersymmetric partition functions as well.

\section{Giant graviton expansion of the multi-matrix model}
\label{sec:gge}

Following \cite{deMelloKoch:2025cec}, we consider a quantum mechanical matrix-vector model with a set of $d$ Hermitian $N\times N$ matrices $\Phi_I$, $I=1,2,\ldots,d$ coupled to $f$ complex vectors $\phi_a$, $\phi_a^\dagger$, $a=1,2,\ldots f$, with the free action
\begin{equation}
    S=\int dt\left[\dot\phi_a^\dagger\dot\phi_a-\phi_a^\dagger\phi_a+\Tr(\dot\Phi_I\dot\Phi_I-\Phi_I\Phi_I)\right].
\end{equation}
We gauge the global $U(N)$ symmetry so that we are only interested in gauge invariant operators that are singlets under $U(N)$.  Such operators can be built out of traces of combinations of $\Phi_I$ as well as bilinears $\phi_a^\dagger(\cdots)\phi_b$ with an arbitrary set of matrices sandwiched in between.

We can write the partition function in terms of fugacities $x_a$, $y_a$ and $q_I$ corresponding to $\phi_a$, $\phi_a^\dagger$ and $\Phi_I$, respectively
\begin{equation}
    Z(x_a,y_a,q_I)=\int_{U(N)}DU\PE\left(f_{\mathrm{adj}}(q_I)\chi_{\mathrm{adj}}(U)+f_F(x_a)\chi_F(U)+f_{\bar F}(y_a)\chi_{\bar F}(U)\right),
\end{equation}
where PE denotes the plethystic exponential, and the single letter `indices' are%
\footnote{Strictly speaking, we are computing a partition function and not an index.  But we will nevertheless refer to these partition functions as indices in order to highlight the similarities with the conventional giant graviton indices.}
\begin{equation}
    f_{\mathrm adj}(q_I)=\sum_Iq_I,\qquad f_F(x_a)=\sum_ax_a,\qquad f_{\bar F}(y_a)=\sum_ay_a,
\end{equation}
and $\chi_{\mathrm{adj}}$, $\chi_F$ and $\chi_{\bar F}$ are $U(N)$ characters in the adjoint, fundamental and anti-fundamental representations, respectively.  Writing out the integral over $U(N)$ in an eigenvalue basis gives us the explicit form for the partition function
\begin{align}
    Z(x_a,y_a,q_I)&=\fft1{N!}\int\prod_i\fft{dz_i}{2\pi iz_i}\prod_{i\ne j}(1-z_{ij})\PE\left(\sum_I\sum_{i,j}q_Iz_{ij}+\sum_a\sum_i(x_az_i+y_az_i^{-1})\right)\nn\\
    &=\fft1{N!}\int\prod_i\fft{dz_i}{2\pi iz_i}\prod_{i\ne j}(1-z_{ij})\prod_I\prod_{i,j}\fft1{(1-q_Iz_{ij})}\prod_a\prod_i\fft1{(1-x_az_i)(1-y_az_i^{-1})}.
\end{align}

Motivated by the giant graviton expansion of super-Yang-Mills partition functions, we seek a similar expansion for this bosonic matrix-vector model.  In order to do so, we restrict ourselves to adjoint representations of the $U(N)$ symmetry group.  In particular, we disregard the complex vectors and focus only on the $U(N)$ multi-matrix model with partition function
\begin{align}
    Z(q_I)&=\fft1{N!}\int\prod_i\fft{dz_i}{2\pi iz_i}\prod_{i\ne j}(1-z_{ij})\PE\left(\sum_I\sum_{i,j}q_Iz_{ij}\right)\nn\\
    &=\fft1{N!}\int\prod_i\fft{dz_i}{2\pi iz_i}\prod_{i\ne j}(1-z_{ij})\prod_I\prod_{i,j}\fft1{(1-q_Iz_{ij})}.
\label{eq:dmmpf}
\end{align}
In general, this is a model of $d$ unitary matrices.  For $d=1$, the partition function for the single-matrix model counts gauge invariant operators of the form $\Tr(X^n)$.  In this case, the bosonic partition function is identical to the 1/2-BPS index of $\mathcal N=4$ SYM, with single letter index $f(q)=q$.  For $d=2$, the corresponding operators take the form $\Tr(X^nY^m)$, and the single letter index is $f(q_I)=q_1+q_2$.  This differs from the $1/4$ BPS index of $\mathcal N=4$ SYM, as the latter only counts symmetrized traces with corresponding single letter index $f_{1/4}(x,y)=x+y-xy$.  Nevertheless, the matrix model constructions of the bosonic partition function and the superconformal index share many common features.  As we now demonstrate, this includes the existence of a giant graviton expansion even in the absence of supersymmetry.

\subsection{Murthy's giant graviton expansion}

While the usual giant graviton expansion is a string theoretic construction involving graviton modes blowing up into wrapped branes, Murthy demonstrated in Ref.~\cite{Murthy:2022ien} that a large class of unitary matrix model partition functions admits a similar expansion. In particular, any function of the form
\begin{equation}
    Z(\vec q\,)=\fft1{N!}\int\prod_i\fft{dz_i}{2\pi iz_i}\prod_{i\ne j}(1-z_{ij})\PE\left(f(\vec q\,)\sum_{i,j}z_{ij}\right),
\end{equation}
can be rewritten as a giant graviton expansion
\begin{equation}
    \fft{Z_N(\vec q\,)}{Z_\infty(\vec q\,)}=1+\sum_{m\ge1}G_N^{(m)}(\vec q\,),
\end{equation}
where \cite{Murthy:2022ien,Eniceicu:2023uvd,Liu:2022olj}
\begin{align}
    G_N^{(m)}(\vec{q}\,)&=\fft1{(m!)^2}\int\prod_i\left[\fft{dz_i}{2\pi iz_i}\fft{dw_i}{2\pi iw_i}\left(\fft{w_i}{z_i}\right)^N\right]\fft{\prod_{i\ne j}(1-z_{ij})(1-w_{ij})}{\prod_{i,j}(1-w_i/z_j)(1-z_i/w_j)}\nn\\
    &\kern10em\times\PE\Biggl(-\hat f(\vec q\,)\sum_{i,j}(z_i-w_i)(z_j^{-1}-w_j^{-1})\Biggr).
\label{eq:Murthygg}
\end{align}
Here $\hat f(\vec q\,)$ is given by
\begin{equation}
    \hat f(\vec q\,)=\fft{f(\vec q\,)}{1-f(\vec q\,)},
\label{eq:fhat}
\end{equation}
and can be thought of as a dual single-letter index.  This expression can be interpreted as a special case of a $U(M|N)$ supergroup partition function specialized to $U(m|m)$ \cite{Eniceicu:2023uvd,Budzik:2025vzr}.

For the $d$-matrix model, since the single letter index is given by $f(q_I)=\sum_Iq_I$, the dual index can be expressed as
\begin{equation}
   \hat f(q_I)=\fft{\sum_Iq_I}{1-\sum_Iq_I}=\sum_{n\ge1}\Bigl(\sum_Iq_I\Bigr)^n=\left(\sum_{n_1,n_2,\ldots,n_d\ge0}\genfrac{(}{)}{0pt}{0}{n_1+n_2+\cdots+n_d}{n_1,n_2,\ldots,n_d}q_1^{n_1}q_2^{n_2}\cdots q_d^{n_d}\right)-1,
\end{equation}
where
\begin{equation}
    \genfrac{(}{)}{0pt}{0}{n_1+n_2+\cdots+n_d}{n_1,n_2,\ldots,n_d}=\fft{(n_1+n_2+\cdots+n_d)!}{n_1!n_2!\cdots n_d!},
\end{equation}
is a multinomial coefficient.  For notational convenience, we will write this as
\begin{equation}
    \hat f(\vec q\,)=\sum_{\vec n}{}'\binom{|n|}{\vec n}\vec q^{\;\vec n}.
\end{equation}
Note that the prime on the sum indicates that the $\vec n=0$ term is removed.

Inserting $\hat f(\vec q\,)$ into (\ref{eq:Murthygg}) and taking the plethystic exponential gives
\begin{align}
    G_N^{(m)}(\vec q\,)&=\fft1{(m!)^2}\int\prod_i\left[\fft{dz_i}{2\pi iz_i}\fft{dw_i}{2\pi iw_i}\left(\fft{w_i}{z_i}\right)^N\right]\fft{\prod_{i\ne j}(1-z_{ij})(1-w_{ij})}{\prod_{i,j}(1-w_i/z_j)(1-z_i/w_j)}\nn\\
    &\kern10em\prod_{\vec n}{}'\prod_{i,j}
    \left[\fft{(1-\vec q^{\;\vec n}z_{ij})(1-\vec q^{\;\vec n}w_{ij})}{(1-\vec q^{\;\vec n}z_i/w_j)(1-\vec q^{\;\vec n}w_i/z_j)}
    \right]^{\binom{|n|}{\vec n}}\nn\\
    &=\fft1{(m!)^2}\int\prod_i\left[\fft{dz_i}{2\pi iz_i}\fft{dw_i}{2\pi iw_i}\left(\fft{w_i}{z_i}\right)^N\right]\prod_{\vec n}\prod_{i,j}
    \left[\fft{(1-\vec q^{\;\vec n}z_{ij})'(1-\vec q^{\;\vec n}w_{ij})'}{(1-\vec q^{\;\vec n}z_i/w_j)(1-\vec q^{\;\vec n}w_i/z_j)}
    \right]^{\binom{|n|}{\vec n}},
\label{eq:GmN}
\end{align}
where the primes in the second expression indicate that the $\vec n=0$ terms with $i=j$ are to be excluded.  Note that $G_N^{(m)}$ only depends on $N$ through the factor $\left(\prod_i w_i/z_i\right)^N$.  To highlight the $N$ dependence, we perform the $w_i$ contour integrals first, taking $|q_I|<|w_i|<|z_i|$.  Poles that lie inside the $w_i$ contour correspond to $w_i=\vec q^{\;\vec k}z_j$ for $\vec k\ne0$.  However, unlike for the 1/2-BPS index \cite{Eniceicu:2023uvd} or the Schur index \cite{Ezroura:2024wmp}, these poles are not restricted to be only single poles; the order of the pole is given by the corresponding multinomial coefficient $\binom{|k|}{\vec k}$.

The complete residue sum is rather complicated.  However, we can highlight an important subclass of poles.  For the multivariate $w_i$ integral, we have to specify a set of $m$ poles corresponding to $w_1,w_2,\ldots,w_m$.  Each one of these poles is associated with some $z_i$, so that, explicitly
\begin{equation}
    w_1=\vec q^{\;\vec k_1}z_{i_1},\quad w_2=\vec q^{\;\vec k_2}z_{i_2},\quad w_3=\vec q^{\;\vec k_3}z_{i_3},\quad\ldots.
\label{eq:wpoles}
\end{equation}
In general, the $i_j$'s do not have to be distinct.  However, if they are, then we have simply
\begin{equation}
    w_1=\vec q^{\;\vec k_1}z_{1},\quad w_2=\vec q^{\;\vec k_2}z_{2},\quad w_3=\vec q^{\;\vec k_3}z_{3},\quad\ldots,
\label{eq:standard}
\end{equation}
up to $m!$ permutations for associating distinct $z_i$'s with $w_j$'s.  We will call this a set of `standard' poles.  For this standard pole choice, the $N$-dependent factor in (\ref{eq:GmN}) becomes
\begin{equation}
    \left(\prod_i\fft{w_i}{z_i}\right)^N=\left(\vec q^{\;\vec k_1+\vec k_2+\cdots}\right)^N.
\end{equation}
This can be identified as a contribution to the giant graviton index $\hat I^{\mathrm{GG}}_{\vec k}$, where $\vec k=\sum_i\vec k_i$.

Even within this standard pole choice, the $w_i$ poles are generically of higher order than simple poles.  In fact, simple poles only arise from the edge case when each $\vec k_i$ has a single non-zero entry.  In this case, the $w_i$ integrals in (\ref{eq:GmN}) are easily performed, and the result is
\begin{equation}
    \left.G_N^{(m)}(\vec q\,)\right|_{w_i=\vec q^{\;\vec k_i}z_i}=\left(\vec q^{\;\vec k}\right)^N\fft1{m!}\int\prod_i\fft{dz_i}{2\pi iz_i}\prod_{\vec n}\prod_{i,j}\left[\fft{(1-\vec q^{\;\vec n}z_{ij})'(1-\vec q^{\;\vec n+\vec k_i-\vec k_j}z_{ij})'}{(1-\vec q^{\;\vec n+\vec k_i}z_{ij})(1-\vec q^{\;\vec n-\vec k_i}z_{ij})'}\right]^{\binom{|n|}{\vec n}}.
\label{eq:mmZ}
\end{equation}
Again, the primes in the numerator indicate that the $\vec n=0$ terms with $i=j$ are removed, while the prime in the denominator indicates that the $\vec n=\vec k_i$ term is removed when $i=j$.  The latter prime just removes the set of $m$ simple poles, leaving only the residue as the result.  Note that, formally, this expression is independent of the order of the $w_i$ poles.  However, it only provides a valid result for simple poles, as the residue from higher order poles will involve derivatives of the integrand in (\ref{eq:GmN}) that is not captured by the expression (\ref{eq:mmZ}).

By taking a plethystic log of the integrand in (\ref{eq:mmZ}), the standard pole contribution to $G_N^{(m)}$ can be rewritten as
\begin{equation}
    \left.G_N^{(m)}(\vec q\,)\right|_{w_i=\vec q^{\;\vec k_i}z_i}=\left(\vec q^{\;\vec k}\right)^N\fft1{m!}\int\prod_i\fft{dz_i}{2\pi iz_i}\PE\left[m-\sum_{i,j}\fft{(1-\vec q^{\;\vec k_i})(1-\vec q^{\;-\vec k_j})}{1-\vec q}z_{ij}\right].
\end{equation}
Note that the factor $m$ in the plethystic exponential takes into account the omitted terms denoted by the primes in (\ref{eq:mmZ}).  Again, this expression is only valid for the case of simple poles.  We now see that this looks like the partition function of a unitary matrix model, which suggests that the giant graviton expansion of the multi-matrix partition function indeed admits a description in terms of a dual matrix model.  To highlight this connection, we make a further specialization to the choice of $w_i$ poles.

The number and type of giant gravitons corresponding to each term of the residue sum for $G_N^{(m)}$ can be identified by the classical prefactor
\begin{equation}
    \left(\vec q^{\;\vec k}\right)^N=\left(q_1^{m_1}q_2^{m_2}\cdots q_d^{m_d}\right)^N,
\end{equation}
with $m_i$ is the $i$-th component of $\vec k$.  Since $\vec k_i\ne0$, the minimum number of giants is obtained by taking all $|k_i|=1$.  Up to permutations, this is achieved by setting
\begin{align}
    \vec k_i&=(1,0,0,\ldots,0),\qquad \mbox{(first $m_1$ terms)};\nn\\
    \vec k_i&=(0,1,0,\ldots,0),\qquad \mbox{(next $m_2$ terms)};\nn\\
    &\ \vdots\nn\\
    \vec k_i&=(0,0,0,\ldots,1),\qquad \mbox{(final $m_d$ terms)},
\label{eq:minchoicek}
\end{align}
or equivalently by taking the poles
\begin{align}
    w_i&=q_1z_i,\qquad \mbox{(first $m_1$ terms)};\nn\\
    w_i&=q_2z_i,\qquad \mbox{(next $m_2$ terms)};\nn\\
    &\ \vdots\nn\\
    w_i&=q_dz_i,\qquad \mbox{(final $m_d$ terms)},
\label{eq:leadingpc}
\end{align}
with $m=m_1+m_2+\cdots+m_d$.  We can, of course, take higher powers of $q_i$ for these poles, but this would give more than the minimum number of giants for a given $m$.  In general, $G_N^{(m)}$ will contribute to the partition function for $m$ or more giant gravitons.

We can identify the pole configuration, (\ref{eq:leadingpc}), as the leading term in the giant graviton index, $\hat I^{\mathrm{GG,std}}_{(\vec m)}[G_N^{(m)}]$.  In order to provide an explicit form of the index, we change the notation to highlight the decomposition $U(m)\supset U(m_1)\times U(m_2)\times\cdots\times U(m_d)$ corresponding to $d$ distinct species of giant gravitons.  In particular, we take
\begin{align}
    z_i\quad&\to\quad z_i^{(I)}\kern12em I=1,\ldots,d,\quad i=1,\ldots m_I;\nn\\
    \vec k_i\quad&\to\quad\vec k^{(I)}=(0,\ldots,0,1,0,\ldots,0)\qquad\mbox{($1$ in the $I^{\mathrm{th}}$ position)}.
\end{align}
Note that we have ordered the $\vec k_i$'s, which introduces the combinatorial factor, $\binom{|m|}{\vec m}$, when computing the matrix partition function.  The result takes the form
\begin{align}
    \hat I^{\mathrm{GG,std}}_{(\vec m)}[G_N^{(m)}]&=\left(\prod_I q_I^{m_I}\right)^N\prod_I\fft1{m_I!}\oint\prod_I\left[\prod_i\fft{dz_i^{(I)}}{2\pi iz_i^{(I)}}\prod_{i,j}(1-z_{ij}^{(I)})\PE\left(f_v^{(I)}\chi_{\mathrm{adj}}^{(I)}\right)\right]\nn\\
    &\kern11em\times\prod_{I\ne J}\PE\left(f_c^{(I,J)}\chi_F^{(I)}\chi_{\bar F}^{(J)}\right),
\label{eq:GGMM}
\end{align}
where the contour integral over the gauge fugacities $z_i$ is taken over the unit circle.  The single letter indices are
\begin{align}
    f_v^{(I)}&=1-\fft{(1-q_I^{-1})(1-q_I)}{1-\sum_Kq_K},\nn\\
    f_c^{(I,J)}&=\fft{(q_I-1)(1-q_J^{-1})}{1-\sum_Kq_K}=\sqrt{\fft{q_I}{q_J}}\fft{(q_I^{1/2}-q_I^{-1/2})(q_J^{1/2}-q_J^{-1/2})}{1-\sum_Kq_K},
\label{eq:letters}
\end{align}
and the $U(m_I)$ characters are given by
\begin{equation}
    \chi_{\mathrm{adj}}^{(I)}=\sum_{i,j}z_{ij}^{(I)},\qquad\chi_F^{(I)}=\sum_iz_i^{(I)},\qquad\chi_{\bar F}^{(I)}=\sum_i(z_i^{(I)})^{-1}.
\end{equation}
Note that the original partition function, (\ref{eq:dmmpf}), corresponds to $d$ independent matrices all transforming in the adjoint of a single global $U(N)$ group.  However, on the giant graviton side, there are $d$ independent unitary groups, corresponding to $d$ distinct types of giants.

Although we have started with a bosonic matrix model, this form of the giant graviton expansion is highly suggestive of the expansion of a supersymmetric index.  With this interpretation, we may consider $f_v^{(I)}$ to be the single letter index for some $U(m_I)$ theory, and $f_c^{(I,J)}$ to be the single letter index for the bifundamentals between $U(m_I)$ and $U(m_J)$.  In fact, for the single matrix model with $d=1$, there is only one fugacity $q$, and we find
\begin{equation}
    \hat I^{\mathrm{GG,std}}_{(m)}[G_N^{(m)}]=q^{mN}\fft1{m!}\int\prod_i\fft{dz_i}{2\pi iz_i}\prod_{i,j}(1-z_{ij})\PE\left(f_v\chi_{\mathrm{adj}}\right),
\end{equation}
where $f_v=q^{-1}$, which matches the giant graviton expansion of the 1/2 BPS index of $\mathcal N=4$ SYM \cite{Gaiotto:2021xce,Murthy:2022ien}.

At this point, it is important to keep in mind that the ``standard contribution'', (\ref{eq:GGMM}), by itself does not yield the complete giant graviton expansion.  For $d$ matrices, the expansion takes the form
\begin{equation}
    \fft{Z_N(q_I)}{Z_\infty(q_I)}=1+\sum_{\vec m\ne0}q_1^{m_1N}\cdots q_d^{m_dN}\hat I^{\mathrm{GG}}_{(\vec m)}(q_I),
\end{equation}
where the full giant graviton index $I^{\mathrm{GG}}_{(\vec m)}(q_I)$ receives contributions from $G_N^{(1)}$ all the way through $G_N^{(m)}$
\begin{equation}
    \hat I^{\mathrm{GG}}_{(\vec m)}=\sum_{n=1}^m\left(\hat I_{(\vec m)}^{\mathrm{GG,std}}[G_N^{(n)}]+\hat I_{(\vec m)}^{\mathrm{GG,ns}}[G_N^{(n)}]\right).
\label{eq:fullI}
\end{equation}
Here $\hat I_{(\vec m)}^{\mathrm{GG,std}}[G_N^{(n)}]$ is the standard contribution to the index from Murthy's $G_N^{(n)}$, while $\hat I_{(\vec m)}^{\mathrm{GG,ns}}[G_N^{(n)}]$ are additional contributions that arise for `non-standard' choices of the $w_i$ poles in (\ref{eq:wpoles}).  Recall that the standard contribution corresponds to taking all the $i_j$'s to be distinct, as in (\ref{eq:standard}), while the non-standard contribution corresponds to everything else.  The leading order term, 
$\hat I_{(\vec m)}^{\mathrm{GG,std}}[G_N^{(m)}]$, corresponds to the minimum choices of the $\vec k_i$ in (\ref{eq:minchoicek}), and is given by the matrix integral of (\ref{eq:GGMM}).  However, this is just one out of many contributions to $\hat I^{\mathrm{GG}}_{(\vec m)}$.

With so many contributions to a single $\hat I^{\mathrm{GG}}_{(\vec m)}$, one can wonder if there is any special role played by $\hat I_{(\vec m)}^{\mathrm{GG,std}}[G_N^{(n)}]$ in (\ref{eq:GGMM}), which has an elegant interpretation as a giant graviton index.  We can provide a partial answer to this question by noting that while (\ref{eq:GGMM}) is written in the form of a quiver matrix model, the single letters $f_v^{(I)}$ and $f_c^{(I,J)}$ in (\ref{eq:letters}) include inverse powers of some fugacities.  This, of course, originally showed up in the supersymmetric giant graviton expansion \cite{Gaiotto:2021xce,Imamura:2021ytr}, where the interpretation is that the giant graviton fluctuations \textit{decrease} the $R$-charge, resulting in fugacities raised to negative powers.

Since the partition function is essentially a sum over states weighted by the fugacities, one generally takes $|q_I|<1$ to guarantee convergence.  However, once positive and negative powers of $q_I$ enter the sum, it is not obvious that the series even converges.  What we have done implicitly in (\ref{eq:GGMM}) is to define the plethystic exponential of a monomial as
\begin{equation}
    \mathrm{PE}(x)\equiv\fft1{1-x},
\label{eq:PEx}
\end{equation}
for any $x\ne1$.  This analytically continues the series definition of the plethystic exponential beyond the radius of convergence, $|x|<1$, of the series.

Analytic continuation of the plethystic exponential in (\ref{eq:GGMM}) ensures that the integrand is well defined.  However, in the analogous superconformal index case, it is well known that if we strictly integrate the gauge fugacities $z_i$ on the unit circle, as specified in (\ref{eq:GGMM}), while assuming $|q_i|<1$, we would get the incorrect answer, in the sense of not matching the gauge theory index.  Instead, the correct giant graviton expansion of the index requires a deformation of the contour, referred to as pole selection in \cite{Imamura:2021ytr}, or, equivalently, an appropriate analytic continuation prescription for computing the reside sum \cite{Gaiotto:2021xce,Lee:2022vig,Imamura:2022aua,Ezroura:2024wmp}.  One way to view this analytic continuation is to formally assume $|x|<1$ for any factor $x$ that is obtained by plethystic exponentiation in (\ref{eq:PEx}).

The advantage of Murthy's approach to the giant graviton expansion \cite{Murthy:2022ien} is that it completely bypasses the pole selection issue by strictly keeping the integration contour of (\ref{eq:GGMM}) on the unit circle.  This removes possible ambiguities in determining what residues to include and what to exclude.  However, this brings up an issue when comparing the traditional giant graviton expansion (with deformed contour) and the matrix integral (\ref{eq:GGMM}) (with undeformed contour) derived from Murthy's expansion.  While the integrand is in principle the same in both cases, since the choice of poles is different, the two results cannot be the same; they must differ by the sum of the residues that are included/excluded by the different contours.

The resolution to this issue lies in the nature of Murthy's expansion, (\ref{eq:fullI}), which can be written as
\begin{equation}
    \hat I^{\mathrm{GG}}_{(\vec m)}=\hat I^{\mathrm{GG,std}}_{(\vec m)}[G_N^{(m)}]+\hat I^{\mathrm{GG,ns}}_{(\vec m)}[G_N^{(m)}]+\sum_{n=1}^{m-1}\left(\hat I_{(\vec m)}^{\mathrm{GG,std}}[G_N^{(n)}]+\hat I_{(\vec m)}^{\mathrm{GG,ns}}[G_N^{(n)}]\right).
\label{eq:bigsum}
\end{equation}
In the traditional giant graviton expansion, $\hat I^{\mathrm{GG}}_{(\vec m)}$ on the left-hand side is given by the same integral as (\ref{eq:GGMM}), \textit{except} that the contour is deformed according to the pole selection criterion. On the other hand, the first term on the right-hand side is simply (\ref{eq:GGMM}) without any contour deformation.  The difference between these two terms are the residues related to inverse powers of fugacities that would be excluded by lying outside the unit circle but are purposely included in the pole selection procedure.  From this point of view, the remaining terms on the right-hand side of (\ref{eq:bigsum}) effectively implement the pole selection procedure by adding in the appropriate residues while maintaining unit circle contours. For example, the first additional standard contribution to $\hat I^{\mathrm{GG}}_{(\vec m)}$ comes from $G_N^{(m-1)}$, which has one fewer contour integral, and this is exactly what we would expect if we were to add in one of these residues pertaining to pole selection.  In this sense, the right-hand side of (\ref{eq:bigsum}) can be thought of as a rigorous definition of the pole selection procedure.  While Murthy's approach requires summing over many terms, the advantage here is that every contour is well defined, and no guesswork is needed on how to deform the contour.

One important feature of Murthy's expansion for the multi-matrix model worth noting is the presence of higher order poles in (\ref{eq:GmN}) related to the multinomial coefficient $\binom{|n|}{\vec n}$.  Such poles do not show up in the expansion of the 1/2 BPS index \cite{Eniceicu:2023uvd} or the Schur index \cite{Ezroura:2024wmp} but can be present in more general cases such as the one here.  Since the residue theorem for higher order poles involves derivatives of the integrand, this will bring down factors of $N$ coming from derivatives of $(w_i/z_i)^N$ in (\ref{eq:GmN}).  While the resulting expansion, (\ref{eq:bigsum}), remains valid with higher order poles, the various contributions will now carry explicit $N$ dependence.  What this implies is that the giant graviton indices $\hat I^{\textrm{GG}}_{(\vec m)}$ will depend on $N$ and hence carry a record of the original global $U(N)$ symmetry group.  This is in contrast to the physical interpretation of the giant graviton index in the super-Yang-Mills case, where the rank of the gauge group only shows up in the classical factor $(\vec q^{\vec m})^N$ in front of $\hat I^{\textrm{GG}}_{(\vec m)}$.

Curiously, applying Murthy's expansion to the 1/16 BPS super-Yang-Mills index will involve double poles \cite{Ezroura:2024wmp}, which would generate terms linear in $N$ in $G_N^{(m)}$.  Unless there is some additional cancellation, this $N$ dependence will show up in the giant graviton index $\hat I^{\mathrm{GG}}_{(m_1,m_2.m_3)}$ starting with $\hat I^{\mathrm{GG}}_{(1,1,1)}$ corresponding to three D3-brane giants with one moving on each of the three orthogonal rotation planes in $\mathbb C^3\supset S^5$.  If this is the case, then it would be in tension with the expansion of \cite{Imamura:2021ytr}.

\section{Example with \texorpdfstring{$d = 2$}{d=2} Matrices}
\label{sec:d=2}

In the previous section, we derived an expansion of the partition function of the multi-matrix model for an arbitrary number, $d$, of matrices. At $d = 1$, we saw that the bosonic partition function matches that of the 1/2 BPS index of $\mathcal{N} = 4$ SYM.  In this case, Murthy's giant graviton expansion reduces to expressions previously derived in the literature. However, the case for $d\ge2$ is new and does not have any obvious superconformal index analog.

As $d=2$ already gives rise to many of the distinct features of the multi-matrix model and avoids the complication of excessive bookkeeping and combinatorics, we now examine the $d=2$ matrix case in detail.  In particular, we will present explicit calculations demonstrating how Murthy's $G^{(m)}_N$ expansion can be reorganized into a brane-like giant graviton expansion in terms of $\hat I^{\mathrm{GG}}_{(n_1,n_2)}$ such that
\begin{equation}
    \frac{Z_N(q_1,q_2)}{Z_\infty(q_1,q_2)} = \sum_{m=0}^\infty G_N^{(m)}(q_1,q_2)=\sum_{n_1,n_2 = 0}^\infty q_1^{n_1 N}q_2^{n_2 N}\hat I^{\mathrm{GG}}_{(n_1,n_2)}(q_1,q_2),
\label{eq:Exp}
\end{equation}
with $\hat I^{\mathrm{GG}}_{(0,0)} = 1$.  Although the procedure for extracting $\hat I^{(\mathrm{GG})}_{(\vec n)}$ from $G_N^{(m)}$ by decomposing the residue sum can be applied to arbitrarily high orders in the expansion, it gets more and more cumbersome as the order increases.  Hence, we only show results for the first two orders in the fugacities, although the core concepts and potential difficulties for general order will be apparent here.

We recall that $\hat I^{\mathrm{GG}}_{(n_1,n_2)}$ receives contributions from all $G^{(m)}_N$ such that $m \leq n_1 + n_2$.  So for one giant, $\hat I^{\mathrm{GG}}_{(0,1)}$ and $\hat I^{\mathrm{GG}}_{(1,0)}$ will receive contributions for $G^{(1)}_N$, and for two giants, $\hat I^{\mathrm{GG}}_{(0,2)}$, $\hat I^{\mathrm{GG}}_{(2,0)}$, and $\hat I^{\mathrm{GG}}_{(1,1)}$ will receive contributions from both $G^{(1)}_N$ and $G^{(2)}_N$.  Therefore, up to two giants, we only need to investigate $G^{(1)}_N$ and $G^{(2)}_N$.  Furthermore, by symmetry, pairs of indices are related according to $\hat I^{\mathrm{GG}}_{(n_1,n_2)}(q_1,q_2)=\hat I^{\mathrm{GG}}_{(n_2,n_1)}(q_2,q_1)$, so we will make use of this symmetry to only display terms with $n_1\ge n_2$. As we have two fugacities $q_1$ and $q_2$,  we will treat them symmetrically and use the parameter $\epsilon$ (with $q_i\sim\epsilon$) to track the total degree of our terms in both fugacities.

\subsection{The \texorpdfstring{$G_N^{(1)}$}{one Murthy giant} contributions}
We begin with $G^{(1)}_N$, which can be read off directly from (\ref{eq:GmN}).  After performing the trivial $z$ integral, this becomes
\begin{equation}
    G_N^{(1)}(\vec q\,)=\int\left[\fft{dw}{2\pi iw}w^N\right]
    \prod_{\vec n}
    \left[\fft{(1-\vec q^{\;\vec n})'^2}{(1-\vec q^{\;\vec n}/w)(1-\vec q^{\;\vec n}w)}
    \right]^{\binom{|n|}{\vec n}},
\label{eq:G1N}
\end{equation}
where the product is over $\vec n\ge\vec0$ and the prime in the numerator indicates that the $\vec n=\vec0$ term is omitted.  Since there is only a single $w$ integral, $G_N^{(1)}$ receives only standard contributions of the form (\ref{eq:standard}) from the residue sum.  Although we will focus on $d=2$ matrices, so far this expression is general for any $d$. We again note that the poles of this integrand inside the contour are at $w = q^{\;\vec n}$ excluding $\vec n=\vec0$ as the $w$ contour is taken slightly inside the unit circle.  These poles will give a prefactor of $q^{\;\vec nN}$ from the $w^N$ term, so the corresponding residues contribute to the brane indices $\hat I^{(\mathrm{GG})}_{(\vec n)}$.  It is important to note that the order of the $\vec q^{\;\vec n}$ pole is given by the multinomial coefficient $\binom{|n|}{\vec n}$; simple poles only occur at the edges when a single $n_i$ is non-zero.  All other choices of $\vec n$ will lead to higher order poles.

Specializing now to $d=2$, we see that the contribution of $G_N^{(1)}$ to $\hat I^{\mathrm{GG}}_{(\ell,0)}$ comes for the simple pole residue at $w=q_1^\ell$
\begin{equation}
    \hat I^{\mathrm{GG}}_{(\ell,0)}[G^{(1)}_N] = 
    \prod_{n_1,n_2\ge0}
    \left[\fft{(1-q_1^{n_1}q_2^{n_2})'^2}{(1-q_1^{n_1-\ell}q_2^{n_2})'(1-q_1^{n_1+\ell}q_2^{n_2})}
    \right]^{\binom{n_1+n_2}{n_1}},
    \label{eq:In0}
\end{equation}
where we have explicitly written out the product $\vec q^{\;\vec n}$ and where the primes indicate that terms that would otherwise vanish are removed.  Explicitly, the numerator product avoids the $(n_1,n_2)=(0,0)$ case and the first denominator term avoids the $(n_1,n_2)=(\ell,0)$ case.  By setting $\ell=1$, we get the complete contribution to the single brane giant graviton index
\begin{equation}
    \hat I^{\mathrm{GG}}_{(1,0)} = 
    \prod_{n_1,n_2\ge0}
    \left[\fft{(1-q_1^{n_1}q_2^{n_2})'^2}{(1-q_1^{n_1-1}q_2^{n_2})'(1-q_1^{n_1+1}q_2^{n_2})}
    \right]^{\binom{n_1+n_2}{n_1}}.
    \label{eq:I10}
\end{equation}
This is complete as it stands, as it cannot receive any additional contributions for $G^{(m)}_N$ with $m\ge2$.  Note, however, that the two giant term
\begin{equation}
    \hat I^{\mathrm{GG}}_{(2,0)}[G^{(1)}_N] = 
    \prod_{n_1,n_2\ge0}
    \left[\fft{(1-q_1^{n_1}q_2^{n_2})'^2}{(1-q_1^{n_1-2}q_2^{n_2})'(1-q_1^{n_1+2}q_2^{n_2})}
    \right]^{\binom{n_1+n_2}{n_1}},
    \label{eq:I20}
\end{equation}
will receive additional contributions for $G^{(2)}_N$ which we will turn to below.

Before looking at $G^{(2)}_N$, however, there is one more residue to consider, and that is the one which arises from the double pole at $w = q_1q_2$.  This yields a contribution
\begin{align}
    \hat I^{\mathrm{GG}}_{(1,1)}[G^{(1)}_N] &=
    \prod_{n_1,n_2\ge0}
    \left[\fft{(1- q_1^{n_1}q_2^{n_2})'^2}{(1-q_1^{n_1-1}q_2^{n_2-1})'(1-q_1^{n_1+1}q_2^{n_2+1})}
    \right]^{\binom{n_1+n_2}{n_1}}\nn\\
    &\quad\times\left[N-1+\sum_{n_1,n_2\ge0}\binom{n_1+n_2}{n_1}\left(\fft1{(1-q_1^{n_1+1}q_2^{n_2+1})}-\fft1{(1-q_1^{n_1-1}q_2^{n_2-1})'}\right)\right].
    \label{eq:I11}
\end{align}
We remark here that this term contains the first $N$-dependent coefficient in our expansion and is a result of the derivative necessary to evaluate the second order pole's residue. In general, we expect $O(N^\ell)$ coefficients for poles of order $\ell +1$.  The index $\hat I^{\mathrm{GG}}_{(1,1)}$ will also receive contributions from $G_N^{(2)}$.

\subsection{The \texorpdfstring{$G_N^{(2)}$}{two Murthy giant} contributions}
Now, we turn to $G^{(2)}_N$, which first contributes at the two giant graviton level.  This will supply the remaining terms in $\hat I^{\mathrm{GG}}_{(0,2)}$, $\hat I^{\mathrm{GG}}_{(2,0)}$, and $\hat I^{\mathrm{GG}}_{(1,1)}$ beyond the $G^{(1)}_N$ contributions given above in (\ref{eq:I20}) and (\ref{eq:I11}).  Setting $m=2$ in (\ref{eq:GmN}) gives the explicit integral expression
\begin{multline}
    G_N^{(2)}(\vec q\,)=\fft1{(2!)^2}\int\prod_i\left[\fft{dz_1dz_2}{2\pi iz_12\pi iz_2}\fft{dw_1dw_2}{2\pi iw_12\pi iw_2}\left(\fft{w_1w_2}{z_1z_2}\right)^N \right] \\
    \prod_{\vec n}
    \bigg[\fft{(1-\vec q^{\;\vec n})^{'4}(1-\vec q^{\;\vec n}z_1/z_2)(1-\vec q^{\;\vec n}z_2/z_1)}{(1-\vec q^{\;\vec n}z_1/w_1)(1-\vec q^{\;\vec n}z_1/w_2)(1-\vec q^{\;\vec n}z_2/w_1)(1-\vec q^{\;\vec n}z_2/w_2)} 
    \\ \times \fft{(1-\vec q^{\;\vec n}w_1/w_2)(1-\vec q^{\;\vec n}w_2/w_1)}{(1-\vec q^{\;\vec n}w_1/z_1)(1-\vec q^{\;\vec n}w_1/z_2)(1-\vec q^{\;\vec n}w_2/z_1)(1-\vec q^{\;\vec n}w_2/z_2)} \bigg] ^{\binom{|n|}{\vec n}}.
    \label{eq:G2N}
\end{multline}
As in the previous section, we first select two $w_i$ poles according to (\ref{eq:wpoles}) and then perform one trivial $z$ integral, leaving a final non-trivial integral to be done.  Selection of the $w_i$ poles can either follow the standard arrangement, (\ref{eq:standard}), or a non-standard one.  The lowest order in $q_i\sim\epsilon$ contribution possible is $\epsilon^{2N}$, since for the $w_i$ poles to be inside the unit circle, they must be proportional to positive powers of $q_1$ or $q_2$. We are interested in these lowest order residues that contribute to $\hat I^{\mathrm{GG}}_{(0,2)}$, $\hat I^{\mathrm{GG}}_{(2,0)}$, and $\hat I^{\mathrm{GG}}_{(1,1)}$. We note all relevant poles are single poles, so all these contributions will remain $N$ independent.

The lowest order $w_i$ poles with $|w_i|<|z_j|$ take the form $w_i=q_jz_k$ where $i,j,k=1,2$. We note that if $w_1 = w_2$, then the resultant residue is zero due to the $1-w_i/w_j$ numerator factors. Hence the remaining non-zero residues are of the form $w_i = q_iz_i$, $w_i = q_1z_i$, or $w_i = q_iz_1$, up to permutations.  The first two cases yield standard contributions to $\hat I^{\mathrm{GG}}_{(1,1)}$ and $\hat I^{\mathrm{GG}}_{(2,0)}$, while the third gives a non-standard contribution.

For the first case, choosing the poles $w_1 = q_1z_1$ and $w_2 = q_1z_2$, and accounting for a $z_1\leftrightarrow z_2$ permutation symmetry, yields a standard contribution to $\hat I^{\mathrm{GG}}_{(2,0)}$ of 
\begin{align}
    \hat I^{\mathrm{GG,std}}_{(2,0)}[G^{(2)}_N] &=\fft1{2}\int\fft{dz}{2\pi iz} \prod_{\vec n}\bigg[\fft{(1-q^{\;\vec n})^{'4}}{(1-\vec q^{\;\vec n}/q_1)'^2(1-\vec q^{\;\vec n}z/q_1)(1-\vec q^{\;\vec n}/zq_1)}\nn\\
    &\kern7em\times \fft{(1-\vec q^{\;\vec n}z)^2(1-\vec q^{\;\vec n}/z)^2}{(1-\vec q^{\;\vec n}q_1)^2(1-\vec q^{\;\vec n}q_1z)(1-\vec q^{\;\vec n}q_1/z)} \bigg] ^{\binom{|n|}{\vec n}},
\label{eq:I20G2}
\end{align}
where we have performed one trivial $z$ integral.  The remaining integral is performed with $z$ on the unit circle and yields a residue sum over the possible poles
\begin{equation}
    \mathrm{i})\quad z=\vec q^{\;\vec n}q_1;\qquad \mathrm{ii})\quad z=\fft1{\vec q^{\;\vec n}q_1}; \qquad \mathrm{iii})\quad z=\fft{\vec q^{\;\vec n}}{q_1};\qquad \mathrm{iv})\quad z=\fft{q_1}{\vec q^{\;\vec n}}.
\label{eq:pchoice}
\end{equation}
The first set of poles are always inside the unit circle,  while the second set are always outside.  For $|n|>1$, the third and fourth set of poles are inside and outside the unit circle, respectively.  However, the $|n|=0$ and $1$ cases require more careful consideration.  These potential poles are located at
\begin{center}
    \begin{tabular}{c|c|c}
         $(n_1,n_2)$&case iii)&case iv)  \\\hline
         $(0,0)$&$1/q_1$&$q_1$\\
         $(1,0)$&$1$&$1$\\
         $(0,1)$&$q_2/q_1$&$q_1/q_2$
    \end{tabular}
\end{center}
One can see that the would-be pole at $z=1$ is not a true pole, as it is canceled by a multiple zero in the numerator.  However the other cases are actual poles.  Following the contour integration procedure, we would include the residue at $z=q_1$ and either the residue at $z=q_1/q_2$ or $z=q_2/q_1$, depending on the magnitude of $q_1/q_2$.  This suggests an ambiguity in the computation of $\hat I^{\mathrm{GG}}_{(2,0)}$.  But, as we will see below, this is removed when combined with the non-standard contribution.

Before looking at the non-standard contribution, we complete the standard contribution case by choosing the residues $w_1 = q_1z_1$ and $w_2 = q_2z_2$ along with the permutations $q_1\leftrightarrow q_2$ and $z_1\leftrightarrow z_2$.  This yields a contribution to $\hat I^{\mathrm{GG}}_{(1,1)}$ of 
\begin{align}
    \hat I^{\mathrm{GG,std}}_{(1,1)}[G^{(2)}_N] &= \int\fft{dz}{2\pi iz}\
    \prod_{\vec n}
    \bigg[\fft{(1-q^{\;\vec n})^{'4}(1-\vec q^{\;\vec n}z)(1-\vec q^{\;\vec n}/z)}{(1-\vec q^{\;\vec n}/q_1)'(1-\vec q^{\;\vec n}z/q_2)(1-\vec q^{\;\vec n}/zq_1)(1-\vec q^{\;\vec n}/q_2)'}\nn \\
    &\kern6em\times\fft{(1-\vec q^{\;\vec n}zq_1/q_2)(1-\vec q^{\;\vec n}q_2/q_1z)}{(1-\vec q^{\;\vec n}q_1)(1-\vec q^{\;\vec n}q_1z)(1-\vec q^{\;\vec n}q_2/z)(1-\vec q^{\;\vec n}q_2)} \bigg] ^{\binom{|n|}{\vec n}}.
\label{eq:I11G2}
\end{align}
The only poles inside the unit circle are at $z=\vec q^{\;\vec n}q_2$.  While the denominator factors suggest additional possible poles, they are canceled by corresponding zeros in the numerator.  In particular, this contour integral does not suffer any ambiguities related to the magnitude of $q_1/q_2$.

The final case we consider is the non-standard pole configuration with $w_1 = q_1z_1$ and $w_2 = q_2z_1$.  Including the combinatorial factor related to swapping $q_1\leftrightarrow q_2$ and $z_1\leftrightarrow z_2$ results in
\begin{align}
    G_N^{(2)*}&=(q_1q_2)^N\int\fft{dz}{2\pi iz}z^N
    \prod_{\vec n}
    \bigg[\fft{(1-q^{\;\vec n})^{'4}(1-\vec q^{\;\vec n}z)(1-\vec q^{\;\vec n}/z)}{(1-\vec q^{\;\vec n}/q_1)'(1-\vec q^{\;\vec n}/q_2)'(1-\vec q^{\;\vec n}/q_1z)(1-\vec q^{\;\vec n}/q_2z)}\nn\\
    &\kern10em\times\fft{(1-\vec q^{\;\vec n}q_1/q_2)(1-\vec q^{\;\vec n}q_2/q_1)}{(1-\vec q^{\;\vec n}q_1)(1-\vec q^{\;\vec n}q_2)(1-\vec q^{\;\vec n}q_1z)(1-\vec q^{\;\vec n}q_2z)} \bigg] ^{\binom{|n|}{\vec n}}.
    \label{eq:I20Res}
\end{align}
The asterisk is to remind us that this is a single non-standard residue of $G_N^{(2)}$.  At the moment, this cannot be assigned to any specific brane index $\hat I^{(GG)}_{(\vec n)}$ since the $z^N$ term in the integrand is yet to be determined. Integrating $z$ on the unit circle will pick up poles at $z=\vec q^{\;\vec n}/q_1$ and $z=\vec q^{\;\vec n}/q_2$.  The poles at $1/q_1$ and $1/q_2$ lie outside the contour and can immediately be discarded.  The double pole at $z=1$ is canceled by a double zero in the numerator, and poles at $\mathcal O(\epsilon)$ and higher will not contribute to the two giant graviton indices.  So, the only poles of concern are at $q_1/q_2$ or $q_2/q_1$.  If $|q_1| < |q_2|$, then the former pole is inside the unit circle, and we get a non-standard contribution to $\hat I^{\mathrm{GG}}_{(2,0)}$ of 
\begin{equation}
    \begin{aligned}
    \hat I^{\mathrm{GG,ns}}_{(2,0)}[G^{(2)*}_N] &=
    \prod_{\vec n}{}
    \bigg[\fft{(1-q^{\;\vec n})^{'4}}{(1-\vec q^{\;\vec n}/q_1)'^2(1-\vec q^{\;\vec n}/q_2)'(1-\vec q^{\;\vec n}q_2/q_1^2)}\\
    &\kern2em\times \fft{(1-\vec q^{\;\vec n}q_1/q_2)^2(1-\vec q^{\;\vec n}q_2/q_1)^2}{(1-\vec q^{\;\vec n}q_1)^2(1-\vec q^{\;\vec n}q_1^2/q_2)(1-\vec q^{\;\vec n}q_2)} \bigg] ^{\binom{|n|}{\vec n}}
    \end{aligned}
    \qquad\mbox{if $|q_1|<|q_2|$}.
\label{eq:I20NS}
\end{equation}
However, if $|q_1|>|q_2|$ then $\hat I^{\mathrm{GG}}_{(0,2)}$ will receive a non-standard contribution instead.

If it were to persist, this dependence on the magnitude of $q_1/q_2$ in the non-standard contribution would point to a failure of the giant graviton expansion, as the matrix model index is certainly well-defined as a power series in the fugacities $q_1$ and $q_2$ without ambiguity.  However, as pointed out above, there is a similar dependence of the standard contribution to $\hat I^{\mathrm{GG}}_{(2,0)}$ in (\ref{eq:I20G2}) on $q_1/q_2$.  A careful examination of the residues of (\ref{eq:I20G2}) at either $q_1/q_2$ or $q_2/q_1$ gives
\begin{equation}
    \hat I^{\mathrm{GG,std}}_{(2,0)}[G_N^{(2)}]\bigg|_{z=q_1/q_2}=-\hat I^{\mathrm{GG,std}}_{(2,0)}[G_N^{(2)}]\bigg|_{z=q_2/q_1}=-\fft12\hat I^{\mathrm{GG,ns}}_{(2,0)}[G^{(2)*}_N].
\label{eq:ambig}
\end{equation}
We can now examine the structure of $\hat I^{\mathrm{GG}}_{(2,0)}[G_N^{(2)}]$.  For $|q_1|<|q_2|$, we note that it will receive i) a standard contribution at $z=q_1/q_2$, ii) a non-standard contribution, and iii) a standard residue sum over poles at $z=\mathcal O(\epsilon)$ and higher.  Using (\ref{eq:ambig}), this takes the form
\begin{align}
    \hat I^{\mathrm{GG}}_{(2,0)}[G_N^{(2)}]&=\hat I^{\mathrm{GG,std}}_{(2,0)}[G_N^{(2)}]\bigg|_{z=q_1/q_2}+\hat I^{\mathrm{GG,ns}}_{(2,0)}[G^{(2)*}_N]+\mbox{standard residue sum}\nn\\
    &=\hat I^{\mathrm{GG,std}}_{(2,0)}[G_N^{(2)}]\bigg|_{z=q_2/q_1}+\mbox{standard residue sum}.
\label{eq:ressum}
\end{align}
On the other hand, if $|q_1|>|q_2|$, we would directly obtain the second line of (\ref{eq:ressum}).  As a result, the two-giant graviton contribution is independent of the ratio $q_1/q_2$ and is indeed unambiguous.  We expect any other ambiguous contributions of this form at higher orders in the giant graviton expansion to likewise be resolved.

\subsection{Assembling the brane indices}

At the single giant level, the giant graviton index $\hat I^{\mathrm{GG}}_{(1,0)}$ is given in (\ref{eq:I10}), and $\hat I^{\mathrm{GG}}_{(0,1)}$ can be obtained  by interchanging $q_1\leftrightarrow q_2$.  At the two giant level, all three quantities, $\hat I^{\mathrm{GG}}_{(0,2)}$, $\hat I^{\mathrm{GG}}_{(2,0)}$ and $\hat I^{\mathrm{GG}}_{(1,1)}$ receive contributions from both $G_N^{(1)}$ and $G_N^{(2)}$.  For $\hat I^{\mathrm{GG}}_{(2,0)}$, it is interesting to note that $\hat I^{\mathrm{GG}}_{(2,0)}[G_N^{(1)}]$ in (\ref{eq:I20}) matches the $z=q_1$ and $z=1/q_1$ residues of the standard $G_N^{(2)}$ contribution, (\ref{eq:I20G2}), in the sense that
\begin{equation}
    \hat I^{\mathrm{GG,std}}_{(2,0)}[G_N^{(2)}]\bigg|_{z=q_1}=-\hat I^{\mathrm{GG,std}}_{(2,0)}[G_N^{(2)}]\bigg|_{z=1/q_1}=-\fft12\hat I^{\mathrm{GG}}_{(2,0)}[G^{(1)}_N].
\label{eq:q1q1id}
\end{equation}
We can now assemble the full two-giant index $\hat I^{\mathrm{GG}}_{(2,0)}$ using (\ref{eq:I20}), (\ref{eq:ressum}) and (\ref{eq:q1q1id})
\begin{align*}
    \hat I^{\mathrm{GG}}_{(2,0)}&=\hat I^{\mathrm{GG}}_{(2,0)}[G_N^{(1)}]+\hat I^{\mathrm{GG}}_{(2,0)}[G_N^{(2)}]\\
    &=2\hat I^{\mathrm{GG,std}}_{(2,0)}[G_N^{(2)}]\bigg|_{z=1/q_1}+\hat I^{\mathrm{GG,std}}_{(2,0)}[G_N^{(2)}]\bigg|_{z=q_2/q_1}+\mbox{standard residue sum}\nn\\
    &=\hat I^{\mathrm{GG,std}}_{(2,0)}[G_N^{(2)}]\bigg|_{z=1/q_1}+\hat I^{\mathrm{GG,std}}_{(2,0)}[G_N^{(2)}]\bigg|_{z=q_2/q_1}\\
    &\kern8em+\left(\mbox{standard residue sum}-\hat I^{\mathrm{GG,std}}_{(2,0)}[G_N^{(2)}]\bigg|_{z=q_1}\right).
\end{align*}
Effectively, this demonstrates that that $\hat I^{\mathrm{GG}}_{(2,0)}$ can be obtained completely from the leading order standard contribution, (\ref{eq:I20G2}), provided we only sum over the third type of poles in (\ref{eq:pchoice}), namely those at $z=(q_1^{-1})\vec q^{\;\vec n}$.  (Note that this fully contains the first type of poles.)  This is basically a rigorous derivation of Imamura's pole selection prescription \cite{Imamura:2021ytr} using Murthy's expansion.  In other words, the leading contributions given by (\ref{eq:leadingpc}) define the giant graviton expansion, while the additional standard and non-standard contributions modify the standard contour integrals to pick out a modified set of residues as needed for a proper giant graviton expansion \cite{Ezroura:2024wmp}.

The index $\hat I^{\mathrm{GG}}_{(2,0)}$ can be obtained from $\hat I^{\mathrm{GG}}_{(2,0)}$ by swapping $q_1\leftrightarrow q_2$.  This leaves $\hat I^{\mathrm{GG}}_{(1,1)}$ which receives contributions (\ref{eq:I11}) from $G_N^{(1)}$ and (\ref{eq:I11G2}) from $G_N^{(2)}$.  While $\hat I^{\mathrm{GG}}_{(1,1)}$ does not suffer from any pole ambiguities, it nevertheless has a linear-$N$ dependent term coming from the residue of the double pole when evaluating $G_N^{(1)}$.  This feature has not shown up in any of the superconformal index expansions, so perhaps the distinction is that supersymmetry prevents such $N$-dependent terms from appearing in the brane indices.

\section{Necessity of \texorpdfstring{$N$}{N} dependence in the auxiliary indices}
\label{sec:Ndepend}

Since the reorganization of Murthy's auxiliary indices, $G_N^{(m)}$, into a classical giant graviton expansion in terms of the brane indices $\hat I^{\mathrm{GG}}_{(n_1,n_2)}$ can be somewhat obscure, we performed numerical checks of the expansion at low orders in $q_i\sim\mathcal O(\epsilon)$.  In the following, we display only the first two nontrivial orders in $\epsilon$ and leave denominators of the form $(q_1-q_2)$ unexpanded. For the expansion to be consistent, these denominators must cancel order by order in the number of giants when we assemble the partition function, as it must take the form of a power series without such denominators that vanish in the unrefined limit $q_1=q_2$.

We start with the single giant graviton terms $\hat I^{\mathrm{GG}}_{(1,0)}$ and $\hat I^{\mathrm{GG}}_{(0,1)}$.  Expanding (\ref{eq:I10}) to $\mathcal O(\epsilon^2)$ gives
\begin{equation}
\hat I^{\mathrm{GG}}_{(1,0)} = \fft1{1-\fft{q_2}{q_1}}\left(-q_1-(q_1^2+q_2^2)+\cdots\right),
\label{eq:G1q1}
\end{equation}
and $\hat I^{\mathrm{GG}}_{(0,1)}$ can be obtained by swapping $q_1\leftrightarrow q_2$.  Note that these functions diverge in the unrefined limit.  Adding the two contributions along with their classical factors gives the total single giant index
\begin{equation}
    \mathcal I_{(1)}=q_1^N\hat I^{\mathrm{GG}}_{(1,0)}+q_2^N\hat I^{\mathrm{GG}}_{(0,1)}=-\chi_{(N+1)}(q_1,q_2)-(\chi_{(N+2)}(q_1,q_2)+(q_1q_2)^2\chi_{(N-2)}(q_1,q_2))+\cdots,
\label{eq:calI1}
\end{equation}
where
\begin{equation}
    \chi_{(\ell)}(q_1,q_2)=\fft{q_1^{\ell+1}-q_2^{\ell+1}}{q_1-q_2}=q_1^\ell+q_1^{\ell-1}q_2+\cdots+q_1q_2^{\ell-1}+q_2^\ell,
\end{equation}
is the $U(2)$ character in the $(\ell+1)$-dimensional representation with charge $\ell$.  We see, in particular, that the $1/(q_1-q_2)$ pole cancels, leaving a well-behaved power series in $q_1$ and $q_2$.

Next, we turn to the two giant graviton terms $\hat I^{\mathrm{GG}}_{(2,0)}$, $\hat I^{\mathrm{GG}}_{(0,2)}$ and $\hat I^{\mathrm{GG}}_{(1,1)}$.  As discussed in section~\ref{sec:d=2}, these receive contributions from both $G_N^{(1)}$ and $G_N^{(2)}$.  We start with $\hat I^{\mathrm{GG}}_{(2,0)}$, which receives two standard contributions:
\begin{align}
  \hat I^{\mathrm{GG}}_{(2,0)}[G_N^{(1)}]& = \fft1{\left(1-\frac{q_2}{q_1}\right)^2
   \left(1-\frac{q_2^2}{q_1^2}\right)} \left(-\frac{q_1^5}{q_2}-\frac{q_1^3}{q_2^2}\left(q_1^4+q_1^2 q_2^2+3q_1 q_2^3 +q_2^4\right)+\cdots
   \right), \nn\\
    \hat I^{\mathrm{GG,std}}_{(2,0)}[G_N^{(2)}]& =\fft1{\left(1-\frac{q_2}{q_1}\right)\left(1-\frac{q_2^2}{q_1^2}\right)}\left(\frac{q_1^5}{q_2}+\fft{q_1^4}{q_2^2}(q_1+q_2)(q_1^2+q_2^2)+\cdots\right),
   \label{eq:G1G2q12}
\end{align}
where we have assumed $|q_1/q_2|<1$.  In this case, we also have a nonstandard contribution:
\begin{equation}
    \hat I^{\mathrm{GG,ns}}_{(2,0)}[G_N^{(2)*}] =\frac1{1-\fft{q_2^2}{q_1^2}}\left(-\fft{q_1^5}{q_2}-\fft{q_1^3}{q_2^2}(q_1+q_2)^2(q_1^2+q_2^2)+\cdots\right).
   \label{eq:G2q12NS}
\end{equation}
When we add these contributions together we get
\begin{align}
    \hat I^{\mathrm{GG}}_{(2,0)} = \frac1{\left(1-\frac{q_2}{q_1}\right)^2
   \left(1-\frac{q_2^2}{q_1^2}\right)}
   \left(-\fft{q_1^3}{q_2}(q_1^2-q_1q^2+q_2^2)-\fft{q_1}{q_2^2}(q_1^6+3q_1^3q_2^3+q_1^2q_2^4+q_2^6)+\cdots\right),
   \label{eq:q12}
\end{align}
and $\hat I^{\mathrm{GG}}_{(0,2)}$ can be obtained from this expression by interchanging $q_1\leftrightarrow q_2$. Note that this has a triple pole in $q_1-q_2$, which must cancel when all the contributions are added together.

Finally, $\hat I^{\mathrm{GG}}_{(1,1)}$ has two standard contributions:
\begin{align}
    \hat I^{\mathrm{GG}}_{(1,1)}[G^{(1)}_N] &= \frac1{\left(1-\frac{q_1}{q_2}\right) \left(1-\fft{q_2}{q_1}\right)}\left(-(N+5) q_1^2 q_2^2-(N+4)q_1q_2 \left( q_1^3+2 q_1^2 q_2 +2 q_1q_2^2 + q_2^3\right)+\cdots\right),\nn\\
   \hat I^{\mathrm{GG,std}}_{(1,1)} [G^{(2)}_N] &=\frac1{\left(1-\frac{q_1}{q_2}\right) \left(1-\frac{q_2}{q_1}\right)}\left(q_1^2 q_2^2+q_1q_2 \left(q_1^3+ q_1^2q_2+ q_1q_2^2+q_2^3\right) +\cdots\right) .
\label{eq:G1G2q}
\end{align}
As indicated above, the contribution from $G_N^{(1)}$ has linear-$N$ dependence because of a double pole residue, while the other auxiliary indices up to this order do not.  (The factorized $N$ plus a constant pattern does not persist at higher orders.)  The two contributions can be summed to yield
\begin{equation}
    \hat I^{\mathrm{GG}}_{(1,1)} = \frac1{\left(1-\frac{q_1}{q_2}\right) \left(1-\fft{q_2}{q_1}\right)}\left(-(N+4) q_1^2 q_2^2-q_1 q_2 (q_1+q_2)\left((N+3)(q_1^2+q_1q_2+q_2^2)+q_1q_2\right)+\cdots\right).
    \label{eq:q1q2}
\end{equation}

The total two giant index is a sum of three terms
\begin{equation}
    \mathcal I_{(2)}=q_1^{2N}\hat I^{\mathrm{GG}}_{(2,0)}+q_1^Nq_2^N\hat I^{\mathrm{GG}}_{(1,1)}+q_2^{2N}\hat I^{\mathrm{GG}}_{(0,2)},
\label{eq:calI2}
\end{equation}
that individually have poles in $q_1-q_2$.  However, since the partition function can be expanded as a power series in the fugacities, these poles must cancel when the individual contributions are added together.  At the single giant level, the leading term in (\ref{eq:G1q1}) is a simple pole
\begin{equation}
    \hat I^{\mathrm{GG}}_{(1,0)}=-\fft{q_1^2}{q_1-q_2}+\cdots,
\end{equation}
which cancels in the combination
\begin{equation}
    \mathcal I_{(1)}=q_1^N\hat I^{\mathrm{GG}}_{(1,0)}+q_2^N\hat I^{\mathrm{GG}}_{(0,1)}=-\fft{q_1^{N+2}}{q_1-q_2}+\fft{q_2^{N+2}}{q_1-q_2}+\cdots=-\fft{q_1^{N+2}-q_2^{N+2}}{q_1-q_2}+\cdots,
\end{equation}
yielding the first term in (\ref{eq:calI1}).  For the two giant case, however, both $\hat I^{\mathrm{GG}}_{(2,0)}$ and $\hat I^{\mathrm{GG}}_{(0,2)}$ have triple poles, so their combination can at most cancel the triple pole, leaving behind a second order pole
\begin{align}
    q_1^{2N}\hat I^{\mathrm{GG}}_{(2,0)}+q_2^{2N}\hat I^{\mathrm{GG}}_{(0,2)}&=-\fft{q_1^{2N+8}(q_1^2-q_1q_2+q_2^2)}{q_1q_2(q_1-q_2)^2(q_1^2-q_2^2)}+\fft{q_2^{2N+8}(q_1^2-q_1q_2+q_2^2)}{q_1q_2(q_1-q_2)^2(q_1^2-q_2^2)}+\cdots,\nn\\
    &=-\fft{q_1^{2N+8}-q_2^{2N+8}}{q_1q_2(q_1-q_2)^2(q_1^2-q_2^2)}(q_1^2-q_1q_2+q_2^2)+\cdots\nn\\
    &=-\fft{(q_1^2)^{N+4}-(q_2^2)^{N+4}}{q_1^2-q_2^2}\fft{q_1^2-q_1q_2+q_2^2}{q_1q_2(q_1-q_2)^2}+\cdots\nn\\
    &\to-(N+4)\fft{q_2^{2(N+3)}}{(q_1-q_2)^2}\qquad\mbox{as}\quad q_1\to q_2.
\end{align}
At the same time, $\hat I^{\mathrm{GG}}_{(1,1)}$ has a second order pole of opposite sign
\begin{align}
    q_1^Nq_2^N\hat I^{\mathrm{GG}}_{(1,1)}&=(N+4)\fft{q_1^{N+4}q_2^{N+4}}{q_1q_2(q_1-q_2)^2}+\cdots\nn\\
    &\to(N+4)\fft{q_2^{2(N+3)}}{(q_1-q_2)^2}\qquad\mbox{as}\quad q_1\to q_2,
\end{align}
and thus the poles cancel in the full two giant contribution, (\ref{eq:calI2}).  Note that a first order pole cannot be present because of the $q_1\leftrightarrow q_2$ interchange symmetry.  While the poles cancel, there is a leftover finite term, which gives the two giant index at leading order
\begin{align}
    \mathcal I_{(2)}&=-\biggl[\fft1{q_2q_2}\chi_{(2N+6)}(q_1,q_2)+2q_1q_2\chi_{(2N+2)}(q_2,q_2)+3(q_1q_2)^3\chi_{(2N-2)}(q_1,q_2)+\cdots\nn\\
    &\qquad+(q_1q_2)^2\chi_{(2N)}(q_1,q_2)+2(q_1q_2)^4\chi_{(2N-4)}(q_1,q_2)+3(q_1q_2)^6\chi_{(2N-8)}(q_2,q_2)+\cdots\biggr],\nn\\
\end{align}
where the series ends when we reach $\chi_{(0)}(q_1,q_2)$.

We emphasize that the presence of the $1/(q_1-q_2)^3$ triple pole in $\hat I^{\mathrm{GG}}_{(2,0)}$ necessitates the linear-$N$ dependence in the mixed term $\hat I^{\mathrm{GG}}_{(1,1)}$, as the combination $q_1^{2N}\hat I^{\mathrm{GG}}_{(2,0)}+q_2^{2N}\hat I^{\mathrm{GG}}_{(0,2)}$ can only cancel one order of the pole.  We believe that this pattern continues at higher orders in the giant graviton expansion, explaining the $N^\ell$ coefficients arising from residues of order-$(\ell+1)$ poles present in higher giant graviton auxiliary indices.  While such $N$-dependence has appeared in the giant graviton expansion of superconformal indices at unrefined points, the fully refined indices have all been independent of $N$ as expected for wrapped brane fluctuations that are insensitive to the rank of the dual gauge theory.  So the presence of $N$-dependent giant graviton expressions for the multi-matrix model appears to be a unique feature of the bosonic theory.

\section{Discussion}
\label{sec:disc}

While much of the work on giant graviton expansions has focused on the brane expansion of superconformal indices, we have shown that such expansions also exist for bosonic matrix models.  We did this by first using Murthy's matrix model expansion, which applies to any unitary matrix integral, and then rearranging the terms into a brane-like giant graviton expansion. In particular, the corresponding giant graviton indices $\hat I^{(\mathrm{GG})}_{\vec m}$ take a form reminiscent of those from holographic theories.  This form hints at a possible ``dual" interpretation to this expansion, even though there is no SUSY and no obvious holographic dual in this theory.

For the bosonic $d$-matrix model we investigated, we introduced one fugacity $q_I$ for each matrix.  The giant graviton expansion involves $d$ types of giants, each with a corresponding single letter index
\begin{equation}
    f_v^{(I)}=1-\fft{(1-q_I^{-1})(1-q_I)}{1-\sum_Kq_K}.
\end{equation}
Note that this has a similar structure to the wrapped D3-brane index
\begin{equation}
    f_{(1,0,0)}=1-\fft{(1-y_1^{-1})(1-p)(1-q)}{(1-y_2)(1-y_3)},
\end{equation}
including the inverse power of the fugacity related to the $I$-th giant graviton. This certainly suggests some sort of brane interpretation for the expansion.  This brane picture is further suggested by the presence of additional bifundamental degrees of freedom, (\ref{eq:letters}), presumably associated with the intersection of the giant graviton branes. 

While our results hint at a brane picture, the structure of the giant graviton expansion of this model is quite intricate, and the question remains whether there really is a dual theory to this bosonic matrix model, or is this arrangement just a mathematical coincidence? One potential issue is that that some of the giant graviton indices have $N$-dependent coefficients, which argues against a standard brane interpretation.  However, the reason for the $N$-dependence is the presence of double and higher order poles in Murthy's integral expression, and this also arises in the case of the standard superconformal indices \cite{Eniceicu:2023uvd}.

One reason for $N$-dependence is that some indices can have poles that merge in the unrefined limit.  This is the case for the Schur index of $\mathcal N=4$ SYM, where the flavored single letter index is
\begin{equation}
    f(x,y)=1-\fft{(1-x)(1-y)}{1-xy},
\end{equation}
which leads to the dual single letter index (\ref{eq:fhat})
\begin{equation}
    \hat f=\fft{x}{1-x}+\fft{y}{1-y}=\sum_{n\ge1}x^n+y^n.
\end{equation}
Plethystically exponentiating this term leads to only single poles.  However, if we work in the unflavored limit, $q=x=y$, we would have instead
\begin{equation}
    \hat f=\sum_{n\ge1}2q^n,
\end{equation}
and the degeneracy $2$ then leads to double poles in the matrix integral \cite{Ezroura:2024wmp}.  This suggests that, in some cases, $N$-dependence can drop out when working with fully refined indices.  Nevertheless, there is still a puzzle with the 1/16 BPS index of $\mathcal N=4$ SYM where the single letter index is
\begin{equation}
    f=1-\fft{(1-y_1)(1-y_2)(1-y_3)}{(1-p)(1-q)}.
\end{equation}
The corresponding dual single letter index is
\begin{equation}
\hat f=\fft{(1-p)(1-q)}{(1-y_1)(1-y_2)(1-y_3)}-1=\sum_{n_1,n_2,n_3\ge0}y_1^{n_1}y_2^{n_2}y_3^{n_3}(1+y_1y_2y_3-p-q)-1,
\end{equation}
where we used $pq=y_1y_2y_3$.  We see that degeneracy 2 shows up for $n_1,n_2,n_3\ge1$; the first place this occurs is the term $2y_1y_2y_3$, which will lead to a double pole contribution to the three giant graviton index $\hat I^{(\mathrm{GG})}_{(1,1,1)}$.  However, any such $N$-dependence would be unexpected from the D3-brane picture and appears to be in tension with Imamura's giant graviton picture \cite{Imamura:2021ytr}.  One way to remove the double poles is to relax the index condition $pq=y_1y_2y_3$.  However, this would not solve the issue of $N$-dependence of the actual index.  This puzzle is currently under investigation.

Finally, perhaps the most fundamental question to come out of this work is: what is the core property of a theory that permits a giant graviton expansion with a holographic brane interpretation? Our results seem to suggest that it is not necessarily the presence of supersymmetry or conformal invariance, but perhaps can be a common feature to all matrix models admitting a large-$N$ limit.

\acknowledgments

This material is based upon work supported by the U.S.~Department of Energy, Office of Science, Office of High Energy Physics, under Award Number DE-SC0026542.  BB was supported by a Leinweber summer fellowship.

\medskip

\noindent\textit{Disclaimer.} This report was prepared as an account of work sponsored by an agency of the United States Government.  Neither the United States Government nor any agency thereof, nor any of their employees, makes any warranty, express or implied, or assumes any legal liability or responsibility for the accuracy, completeness, or usefulness of any information, apparatus, product, or process disclosed, or represents that its use would not infringe privately owned rights.  Reference herein to any specific commercial product, process, or service by trade name, trademark, manufacturer, or otherwise does not necessarily constitute or imply its endorsement, recommendation, or favoring by the United States Government or any agency thereof.  The views and opinions of authors expressed herein do not necessarily state or reflect those of the United States Government or any agency thereof.

\bibliographystyle{JHEP}
\bibliography{refs}

\end{document}